\documentclass[conference]{IEEEtran}

\ifCLASSOPTIONcompsoc
  \usepackage[nocompress]{cite}
\else
  \usepackage{cite}
\fi

\ifCLASSINFOpdf
\usepackage[pdftex]{graphicx}
  \graphicspath{{figures/pdf/}{figures/jpeg/}}
  \DeclareGraphicsExtensions{.pdf,.jpeg,.png}
\else
  \usepackage[dvips]{graphicx}
  \DeclareGraphicsExtensions{.eps}
\fi

\usepackage{amsmath}

\ifCLASSOPTIONcompsoc
  \usepackage[caption=false,font=footnotesize,labelfont=sf,textfont=sf]{subfig}
\else
  \usepackage[caption=false,font=footnotesize]{subfig}
\fi

\usepackage{paralist}
\usepackage{graphicx}
\providecommand{\abs}[1]{\lvert#1\rvert}

\begin{document}
\title{Angry or Climbing Stairs? Towards Physiological Emotion Recognition in the Wild}

\author{\IEEEauthorblockN{Judith S. Heinisch, Christoph Anderson, and Klaus David}
\IEEEauthorblockA{Chair for Communication Technology (ComTec)\\
University of Kassel\\
Kassel, Germany\\
Email: comtec@uni-kassel.de}}

\maketitle

\begin{abstract}
Inferring emotions from physiological signals has gained much traction in the last years. Physiological responses to emotions, however, are commonly interfered and overlapped by physical activities, posing a challenge towards emotion recognition in the wild. In this paper, we address this challenge by investigating new features and machine-learning models for emotion recognition, non-sensitive to physical-based interferences. We recorded physiological signals from 18 participants that were exposed to emotions before and while performing physical activities to assess the performance of non-sensitive emotion recognition models. We trained models with the least exhaustive physical activity (sitting) and tested with the remaining, more exhausting activities. For three different emotion categories, we achieve classification accuracies ranging from 47.88\% - 73.35\% for selected feature sets and per participant. Furthermore, we investigate the performance across all participants and of each activity individually. In this regard, we achieve similar results, between 55.17\% and 67.41\%, indicating the viability of emotion recognition models not being influenced by single physical activities.
\end{abstract}


\IEEEpeerreviewmaketitle

\section{Introduction}
Sensing and understanding emotional states of individuals are one of the main challenges in the field of Human-Computer-Interaction. Research in this field is fueled by the idea to enhance computer systems to a state where they can sense, adapt, or even react to emotional states of their users. For example, advanced driver assistant systems might sense emotional states of drivers to detect risky driving behaviors~\cite{Raja:2017}. Work-related environments may include emotion recognition to support software developers in their productivity and mitigate effects caused by interruptions~\cite{Fritz:2016}. Physical and physiological responses to emotions have been investigated to facilitate various applications of emotion recognition~\cite{Shu:2018}. Among others, microphones and cameras have been used to extract speech, facial expressions, or postures for physical-based emotion recognition~\cite{Jerritta:2011}. Physical responses, however, are subject to suppression and disguise as individuals can control facial expressions or the tune of their speech, therefore, confounding emotion recognition systems~\cite{Kim:2004,Gunes:2010}. Contrary to physical signals, physiological responses to emotions cannot easily be triggered and controlled by individuals, but instead, are interfered by physical movement and activity~\cite{Jerritta:2011}.

Approaches that cope with physical-based interferences, for example, provide models designated and trimmed for individual activities~\cite{Hong:2012}, or select appropriate machine-learning models for similar interferences~\cite{Ramos:2014}. Although these approaches are exceedingly practical, they still depend on the type of interference or increase the computation complexity as multiple models are required. Therefore, we aim to address the challenge of recognizing emotions~\textit{throughout} and~\textit{non-sensitive} to physical activities. To investigate physical-based interferences, we carried out an experiment with $18$ participants where emotions were elicited while performing physical activities. To force non-sensitivity, we first filtered the recorded physiological signals and then trained three different machine-learning algorithms with the least exhaustive activity. More exhausting activities were then used to evaluate and to assess the performances of machine-learning models. We found that features based on the linear regression line of physiological signals facilitate machine-learning models that reasonably distinguish between three different categories of emotions. 
The contributions of our paper are four-fold:
\begin{itemize}
  \setlength{\itemsep}{0.25\baselineskip}
  \item We present results of our experiment with $18$ participants.
  \item We investigate the influence of five physical activities on physiological-based emotion recognition.
  \item We present features to recognize emotions during physical activities.
  \item We publish the data set from our experiment, encouraging researchers to work in the field of emotion recognition\footnote{The link to the data will be inserted after acceptance}.
\end{itemize}
The rest of the paper is organized as follows: In Section~\ref{related_work}, we present the state of the art of emotion recognition, focusing on systems that utilize physiological responses to emotion and stress. In Section~\ref{hypothesis}, we derive research questions based on the related work, setting the aim of this paper. In Section~\ref{experimental_setup}, we outline the underlying emotion model as well as emotion categories and describe the setup of our experiment. In Section~\ref{methodology} and~\ref{evaluation}, we detail our approach, elaborating on preprocessing and features that facilitate emotion recognition during physical activities. Finally, we discuss the results of this research before we conclude our paper.
\section{Related Work}
\label{related_work}
The number of wearables already embedding physiological sensors is continuously rising~\cite{Statista:2017}. On the one hand, the pervasiveness of such devices increases the amount of physiological data, covering various facets of our everyday life. Also the fact, that physiological signals cannot be easily suppressed and controlled by individuals, compared to emotion recognition via gestures or facial expressions~\cite{Gunes:2010}, foster the research interest in those wearables. On the other hand, physiological sensors introduce new challenges to the field of emotion recognition. For example, challenges include environmental influences, ambient temperature changes, physical activities, or the consumption of caffeine, sugar, and other non-emotional factors~\cite{Picard:2001}. 

Previous research has already investigated influences of physiological sensors in the field of emotion recognition~\cite{Xu.2017,Picard:2001,Heinisch:2018}. For example, Picard et al. found that physiological signals of one person variate from day-to-day~\cite{Picard:2001}. Furthermore, they found that this day dependence could be handled by applying Sequential Floating Forward Search followed by Fisher Projection. This method led to an accuracy of $81\%$ for classifying eight emotions of one participant over $20$ different days.
Xu et al. investigated the after-effects of physical activities on emotion recognition with physiological signals~\cite{Xu.2017}. Classification accuracies of approximately $20\%$ were achieved with models trained on unaffected data sets when testing on data containing after-effects. To improve the overall classification accuracy, Heinisch et al. merged the aforementioned data sets and applied a selection of commonly used features for emotion recognition~\cite{Heinisch:2018}. They achieved classification accuracies of up to $96\%$. The influence of physical activities on physiological signals has also been investigated in the field of stress detection~\cite{Alamudun:2012,Hong:2012,Ramos:2014}. In~\cite{Alamudun:2012}, Alamudun et al. studied the subject dependence and the influence of activities on stress recognition. By leaving one activity for each participant out, they reached a mean classification accuracy of $66\%$ over $14$ participants and four activities. Hong et al. found an accuracy decrease of $14\%$ training with physically non-interfered stress data and testing with data influenced by exhausting activities~\cite{Hong:2012}. To investigate physical responses to stressors in multiple stimuli scenarios, Hong et al. proposed the use of a two-stage classification for stress recognition. Based on the classified activity (first-stage), a corresponding stress recognition model was applied (second-stage). They achieved a mean classification accuracy of around $88\%$ over $20$ participants.
Ramos et al. improved the two-stage classification proposed in~\cite{Hong:2012} to handle the influence of physical activities on stress detection~\cite{Ramos:2014}. They modified the first-stage by introducing a clustering algorithm. They further trained activity independent models with the clustered data. With this approach, they achieved an accuracy of $65\%$, which was lower than the two-stage method of~\cite{Hong:2012}, but was independent of the kind of physical activity.

Motivated by these approaches, this paper aims to address the influences of physical activities on physiological-based emotion recognition. As there is a vast amount of physical activities we might perform during the day, we believe that emotion recognition models independent to activities still remain an issue. The influence of physical activities on stress detection has been already successful addressed by~\cite{Hong:2012,Ramos:2014}. However, there is still a dependency on physical activities, established by the creation of separated stress detection models in the second stage. This creation might increase the overall effort and complexity of classification models. The same effect is involved in training an emotion classification model with emotion data influenced by a range of different physical activities. In the light of the results of Picard et al.\cite{Picard:2001}, there are still open questions about the significance and generality of different features on emotion recognition.
\section{Goals \& Hypotheses}
\label{hypothesis}
The human body, more precisely, a human's physiological signals are influenced through many factors such as the environment or physical activities~\cite{Picard:2001}. For robust and efficient emotion recognition, we believe that models have to cope with interferences caused, for example, by physical activities. We refer to the term~\textit{non-sensitive} models when pointing towards the ability to cope with interferences. Motivated by existing approaches and studies that already focus on physiological-based emotion recognition, we stress the following research questions:
\begin{itemize}
    \item Can emotion recognition models be trained non-sensitive to physiological interferences (\textit{RQ1}).
    \item Are non-sensitive emotion recognition models robust or are they subject-dependent and susceptible to segmentation parameters (\textit{RQ2}).
\end{itemize}

In the next section, we present the underlying emotion model as well as detail the setup and scenarios of our experiment.
\section{Emotion Model \& Experimental Setup}
\label{experimental_setup}
In our experiment, we used the emotion model by Mehrabian and Russel to categorize emotions~\cite{Mehrabian:1974}. This three-dimensional model classifies emotions in the dimensions of pleasure, arousal, and dominance. Furthermore, we used the International Affective Digital Sounds System (IADS)~\cite{Bradley:2007} to elicit emotions while performing physical activities. Among others, the system contains sounds that relate to the following emotion categories: 

\begin{itemize}
  \item High Positive Pleasure High Arousal (\textit{HPHA}) \\ \textasteriskcentered  pleasure: $6.06 - 7.9$, arousal: $6 - 7.54$
  \item High Negative Pleasure High Arousal (\textit{HNHA}) \\ \textasteriskcentered pleasure: $1.57 - 2.92$, arousal: $6.07 - 8.16$
  \item Neutral (\textit{NEUTRAL}) \\ \textasteriskcentered pleasure: $4.18 - 5.64$, arousal: $4.6 - 5.48$
\end{itemize}

The numbers that are given for each category refer to the rating of sound samples in Self-Assessment Manikin-Scale~\cite{Bradley:1994}. Physiological measurements were recorded using the biosignalsplux toolkit~\cite{Plux:2018} and an E4-wristband~\cite{E3:2014}. Furthermore, we employed smartphone embedded acceleration, gyroscope, gravity, and orientation sensors to record data about a participant's physical activities: sitting, standing, walking, walking upstairs, walking downstairs. For this, we placed the smartphone inside a participant's pocket. We used the same locations for the physiological sensors of the biosignalsplux toolkit as in our previous study~\cite{Heinisch:2018}. The E4-wristband was located on the non-dominant hand and was used to gather a participant's Skin Temperature (ST), the movement with a three-axis acceleration sensor, the Electrodermal Activity (EDA), and the Blood Volume Pulse (BVP). We recorded data from $21$ healthy participants - $11$ female and $10$ male, within the age between $19$ and $50$. The data of three participants were omitted due to erroneous and missing physiological signals resulting in $\approx 300$ minutes of physiological data in total.

\subsection{Scenarios}
\begin{figure}[ht]
    \center
    \resizebox{\columnwidth}{!}{
        \includegraphics{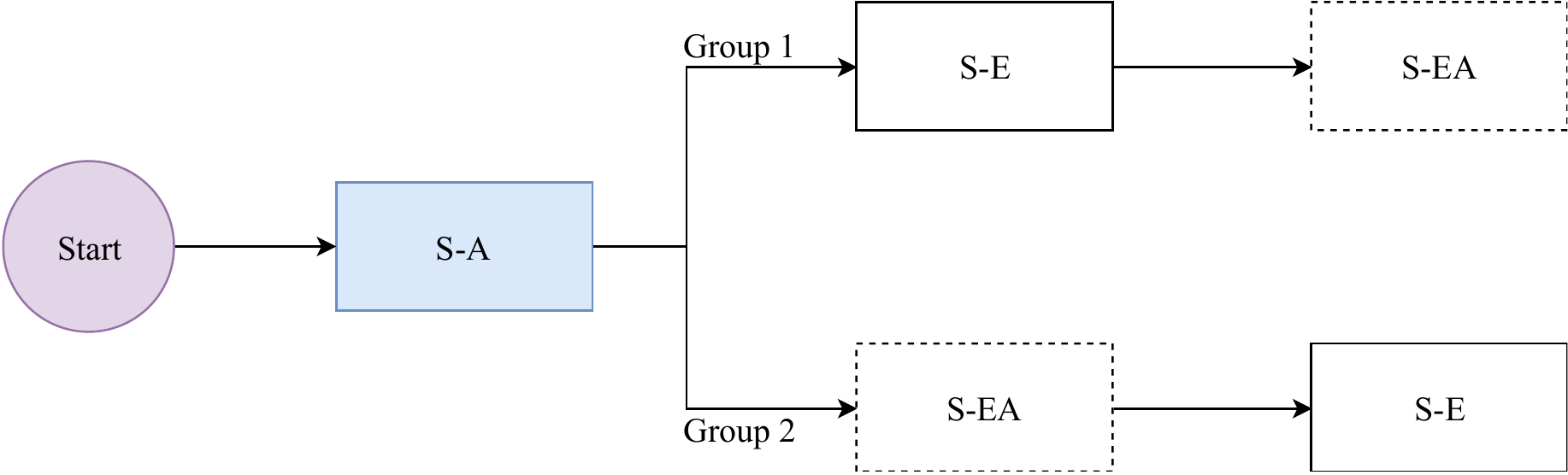}}
    \caption{Scenario procedures for different groups of participants: Scenario Activity (S-A), Scenario Emotion (S-E), Scenario Emotion with Activity (S-EA)}
    \label{fig_scenarios}
\end{figure}

To reduce potential bias, we divided the participants into two groups. Participants from both groups started with the Scenario Activity~(\textit{S-A}) continuing either with the Scenario Emotion~(\textit{S-E}) or the Scenario Emotion with Activity~(\textit{S-EA}) before completing the study with the remaining scenario respectively. Each participant was measured individually. Fig.~\ref{fig_scenarios} details the procedure of the considered scenarios.

\subsubsection*{Scenario Activity (S-A)}
The participant was asked to perform physical activities without any elicitation of emotions. The scenario started with three minutes of resting. After that, the participant performed physical activities, i.e., sitting, standing, walking, walking downstairs and walking upstairs, each for a period of approximately $20$ seconds.

\subsubsection*{Scenario Emotion (S-E)}
In this scenario, the participant sat in a quiet environment, listening to the sounds of each emotion category via headphones to prevent environmental interferences. For each considered emotion category, we chose sound samples for a total period of $2$ minutes. We started with NEUTRAL sound samples. Then, we played the sounds of the HPHA category. To neutralize the influence of the HPHA sounds, we played NEUTRAL sound samples again, before playing HNHA sounds. Finally, NEUTRAL sound samples were played again.

\subsubsection*{Scenario Emotion with Activity (S-EA)}
Finally, we combined both scenarios where emotions were elicited~\textit{while} a participant was performing physical activities. Each participant was asked to perform physical activities in the same order and time as in scenario~\textit{S-A} without resting but while listening to the sounds of one emotion category for each trail. The emotions were the same as in scenario~\textit{S-E}: first NEUTRAL, followed by HPHA and finally HNHA. After each trial of the full set of physical activities, the participant was sitting on a chair and listening to the sound samples related to NEUTRAL again to neutralize the participants' emotional state.

\section{Methodology}
\label{methodology}
In this section, we present the steps towards non-sensitive emotion recognition models. We elaborate on preprocessing and filtering techniques as well as describe the features used in our evaluation.

\subsection{Data Preprocessing}
Different kinds of noise (e.g., caused by moving cables or gaps between the skin and the electrodes) were observed in biosignalsplux sensor data. To reduce the noise, we applied several filtering techniques for each physiological sensor, depict in Table~\ref{table:filtering}. The Electromyogram signal (EMG) was filtered in two different ways. First, we filtered the signal with a fifth-order high-pass Butterworth filter with a cut-off frequency of $40Hz$ - EMG (H) in Table~\ref{table:filtering}. Second, we used a fourth-order low-pass Butterworth filter with a cut-off frequency of $5 Hz$ on the raw signal - EMG (L) in Table~\ref{table:filtering}.
Furthermore, a fourth-order low-pass Butterworth filter with a cut-off frequency of $0.5 Hz$ and $0.25 Hz$ was used to filter the EDA and ST signals, respectively. Before we filtered the Piezoelectric Respiration signal (PZT), a roll median function was used. Then, we filtered the PZT signal with a first-order low-pass Butterworth filter and a cut-off frequency of $1 Hz$. Finally, the PZT signal was normalized. We decided not to filter the E4-wristband signal, as no significant noise was observed.

\begin{table}[ht]
  \caption{Filtering Techniques applied on biosignalsplux data}
  \centering
  \label{table:filtering}
  \begin{tabular}{lll}
    \hline
    \multicolumn{1}{|l|}{\textbf{Sensor}} & \multicolumn{1}{l|}{\textbf{Filtering}}                  & \multicolumn{1}{l|}{\textbf{Units}} \\ \hline
    \multicolumn{1}{|l|}{EMG (H)}         & \multicolumn{1}{l|}{High-pass filter (40Hz, 5th order)}  & \multicolumn{1}{l|}{Micro Volt}     \\ \hline
    \multicolumn{1}{|l|}{EMG (L)}         & \multicolumn{1}{l|}{Low-pass filter (5Hz, 4th order)}    & \multicolumn{1}{l|}{Micro Volt}     \\ \hline
    \multicolumn{1}{|l|}{EDA}             & \multicolumn{1}{l|}{Low-pass filter (0.5Hz, 4th order)}  & \multicolumn{1}{l|}{Micro Siemens}  \\ \hline
    \multicolumn{1}{|l|}{ST}              & \multicolumn{1}{l|}{Low-pass filter (0.25Hz, 4th order)} & \multicolumn{1}{l|}{Celsius}        \\ \hline
    \multicolumn{1}{|l|}{PZT}             & \multicolumn{1}{l|}{\begin{tabular}[c]{@{}l@{}}Rollmedian (7 values, extend),\\ Low-pass filter (1Hz, 1st order),\\ Normalization\end{tabular}}              & \multicolumn{1}{l|}{Percentage}     \\ \hline
  \end{tabular}
\end{table}

\subsection{Window Size}
To assess the robustness of non-sensitive emotion recognition models, we also wanted to investigate the influence of the segmentation parameters, especially of the window size. For our analysis, we used the sliding window algorithm to segment our sensor data and analyzed the influence of different window lengths. We increased the window lengths from $100ms$ to $600ms$ in $50ms$ steps and evaluated the data.

\subsection{Features}
For our evaluation, we used $15$ statistical features on each physiological signal (e.g., mean, standard deviation or the mean of the absolute value of the first difference)~\cite{Picard:1998}. As we have seen in our last paper, the slope of the linear regression line was able to distinguish between different emotion categories for ST~\cite{Heinisch:2018}. Therefore, we further investigated features based on the linear regression line in this paper. After a preliminary analysis of the signals, we found, that some features and sensors were more relevant for the classification than others. Therefore, we evaluated the performance of the models with a second set of features, namely: the mean of the absolute values of the first differences, the absolute value of the slope of the linear regression line, the square root of the absolute value of the intercept of the linear regression line, and the third power of the square root of the absolute value of the intercept of the linear regression line. 

For calculating the linear regression line, we used SciPy, an open-source mathematics library for Python\cite{SciPy:2001}. The $linregress$ function takes two measures and calculates a linear least-squares regression. For our evaluation, we further processed the slope of the regression line, as well as its intercept. Let $W = (x_1, x_2,...x_n)$ be a window with length of $n$ and $I = (1, 2,...n)$ the corresponding $index$ of the elements in $W$. The features are then defined as
\begin{align}
    \begin{split}\label{eq:1}
	    f_{slope} &= \sqrt{\abs{slope(linregress(I,W))}}
	\end{split}\\
	\begin{split}\label{eq:2}
	    f_{intercept}  & = {\sqrt{\abs{intercept(linregress(I,W)}}}   
	\end{split}\\
	\begin{split}\label{eq:3}
	    f'_{intercept} & = {\sqrt{\abs{intercept(linregress(I,W)}}}^3 
	\end{split}
\end{align}
These selected features were calculated on the BVP of the E4-wristband, the ST of the E4-wristband, the EMG (H) and the EMG (L) signals of the biosignalsplux toolkit. In our analysis, we found that the ST, the EMG, and the BVP were useful for classifying the three emotion categories.
\section{Evaluation}
\label{evaluation}
This section describes and compares the results of the evaluation. To investigate the first research question, we trained our models with the physiological signals influenced by the least exhaustive activity (\textit{S-E}) and tested with data influenced by more exhaustive activities (\textit{S-EA}). Then, we separated all activities from scenario \textit{S-EA} and used each activity in the testing phase to evaluate our classifiers empirically. Finally, we investigated the impacts of different window lengths on the classification performance. For the classification, we chose the three best classifiers from our previous research, namely Decision Tree (DT), Random Forest (RF) and K-Nearest Neighbor (KNN, with k=3)~\cite{Heinisch:2018}. For each participant, the classification was done $10$ times for all classifiers. 

\begin{figure}[ht!]
    \centering
    \resizebox{\columnwidth}{!}{
    \includegraphics{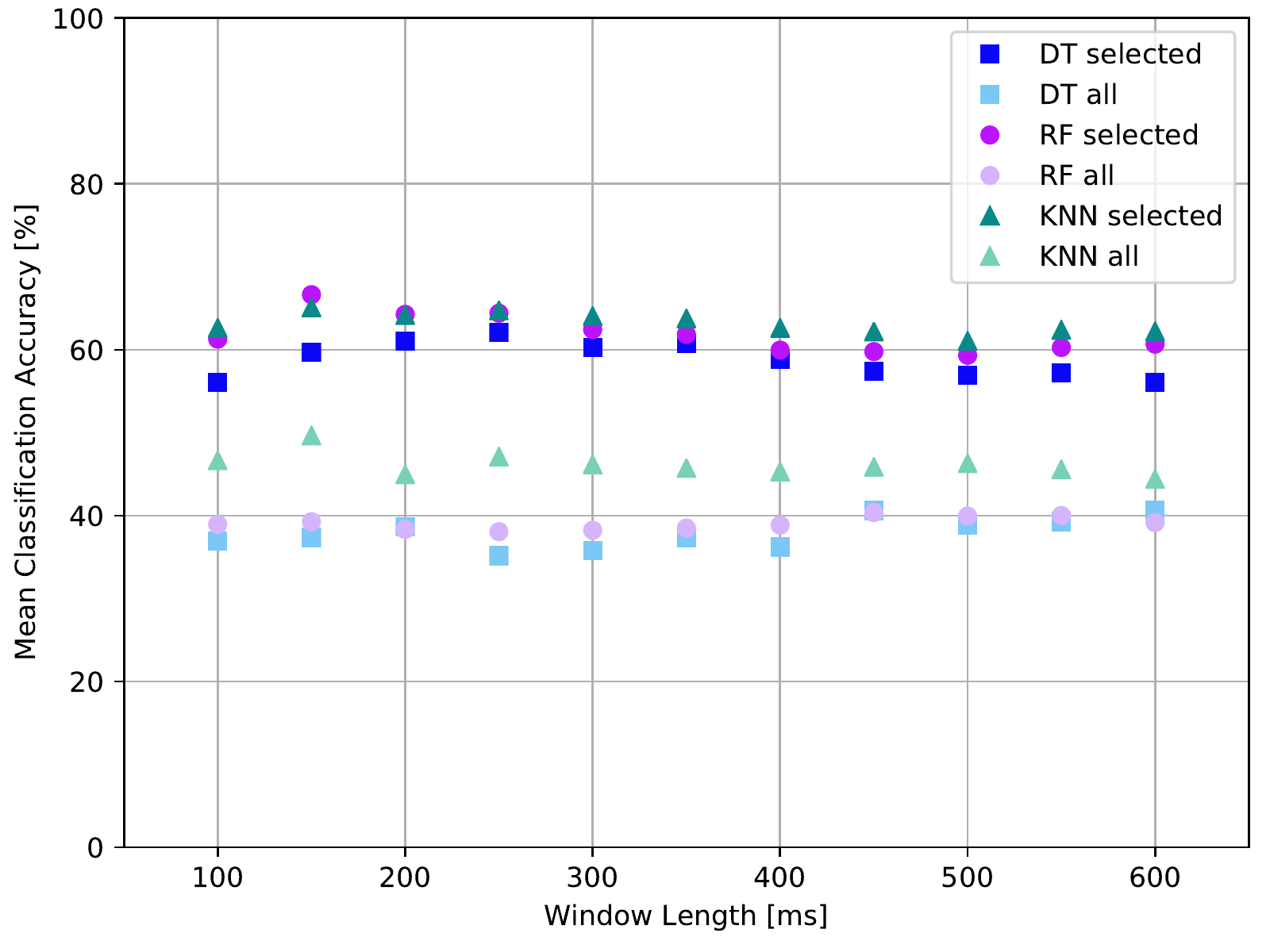}
    }
    \caption{Mean classification accuracy over all participants with different feature sets}
    \label{fig:mean_class_acc}
\end{figure}

\begin{figure}[ht!]
    \centering
    \resizebox{\columnwidth}{!}{
    \includegraphics{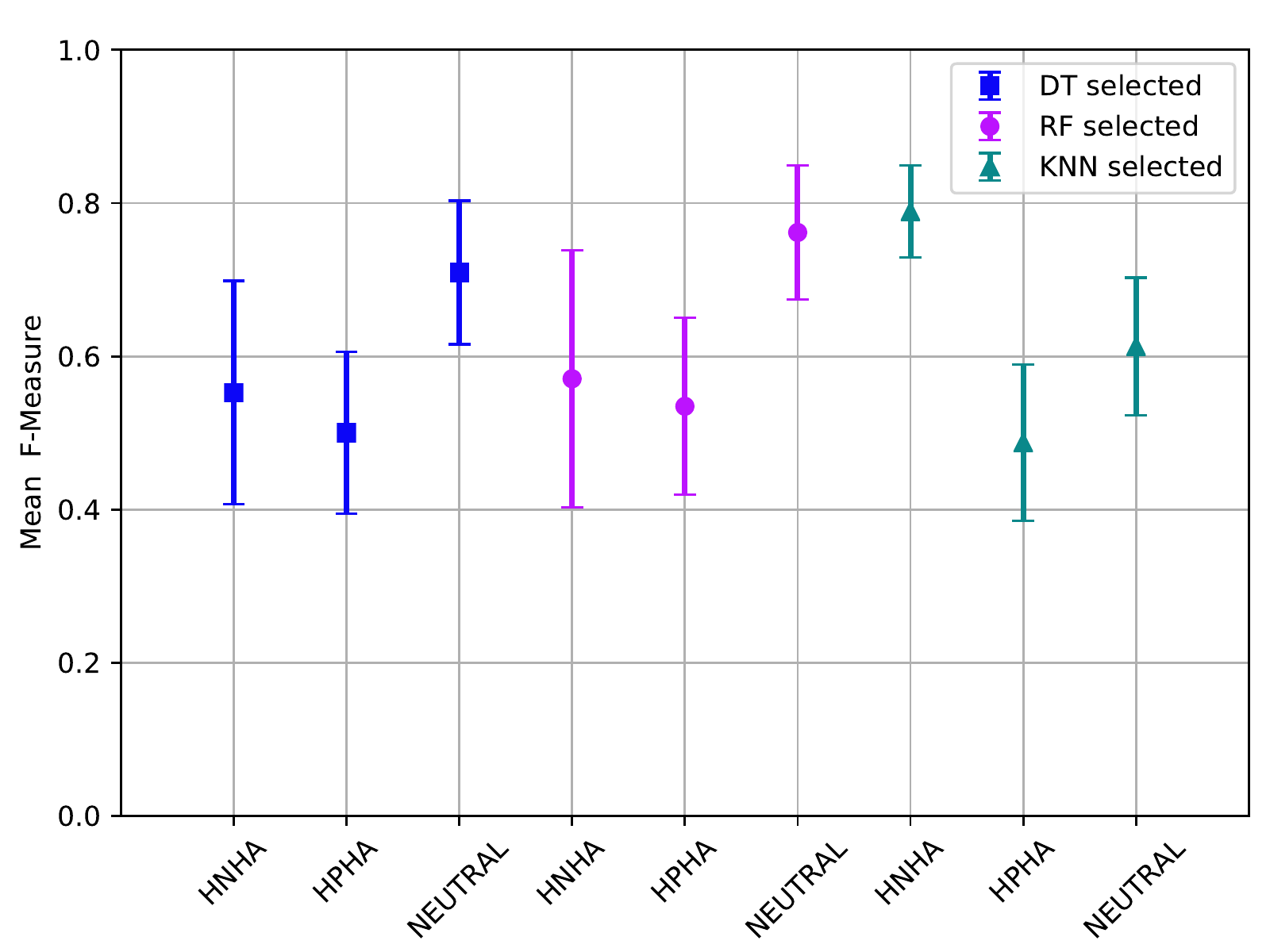}}
    \caption{Mean f-measure over all participants and window sizes using the selected feature set}
    \label{fig:mean_fmeasure}
\end{figure}

Fig.~\ref{fig:mean_class_acc} depicts the mean classification accuracy across all participants and for all activities of~\textit{S-EA}. The KNN classifier achieved the best accuracy over all window sizes. Rather than using all features, where the mean classification accuracy is ranging from $35.17\%$ - $49.64\%$ for all classifiers, the set containing only a selection of features yielded in higher classification accuracy ranging from $56.09\%$ - $66.65\%$. Also the DT, as well as the RF classifier, achieved similar results.

\begin{figure*}[t]
    \centering
    \subfloat[K-Nearest Neighbors (KNN)\label{fig:std_knn}]{%
        \includegraphics[width=0.32\linewidth]{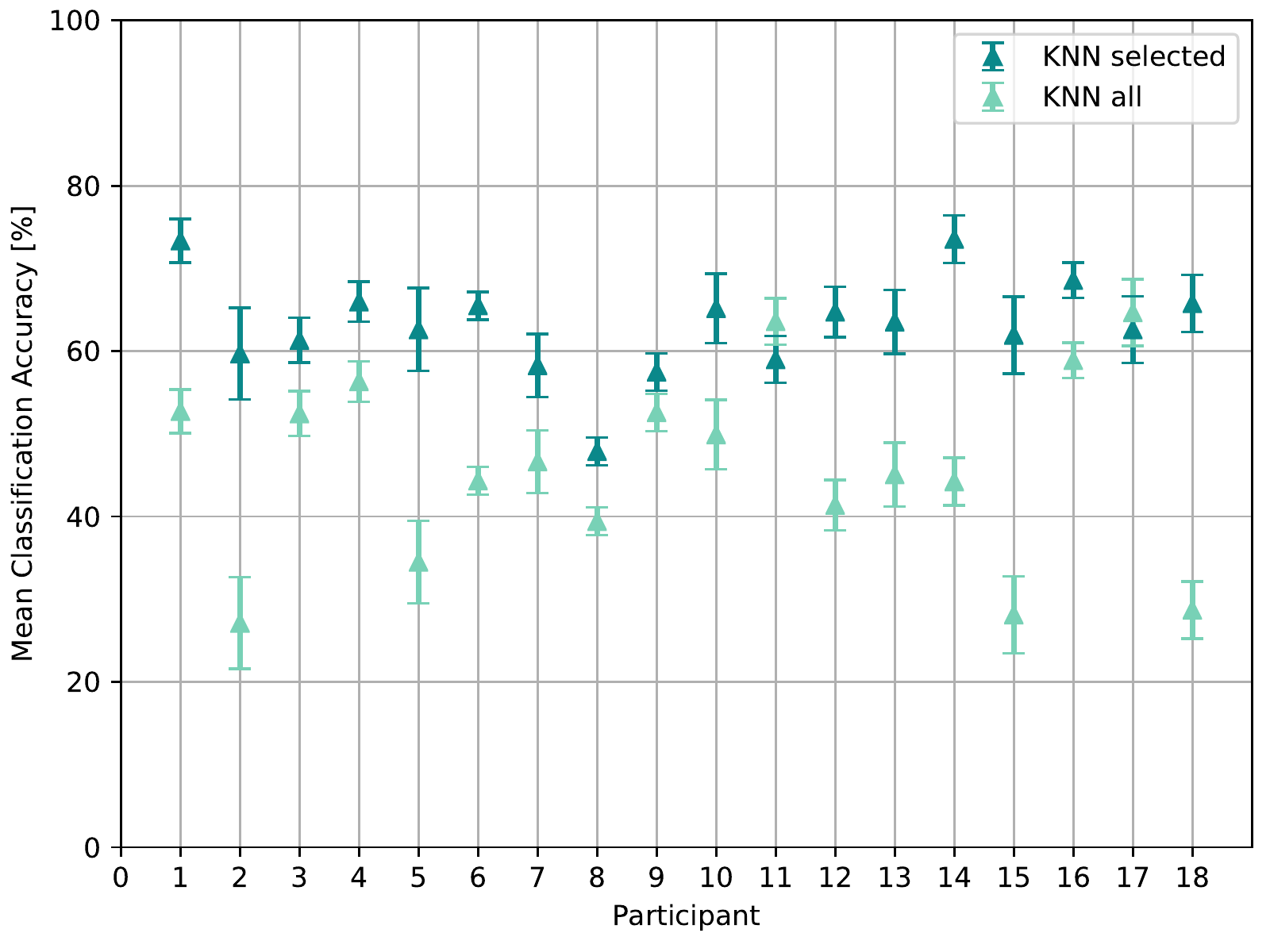}}
    \hfill
    \subfloat[Random Forest (RF)\label{fig:std_rf}]{%
        \includegraphics[width=0.32\linewidth]{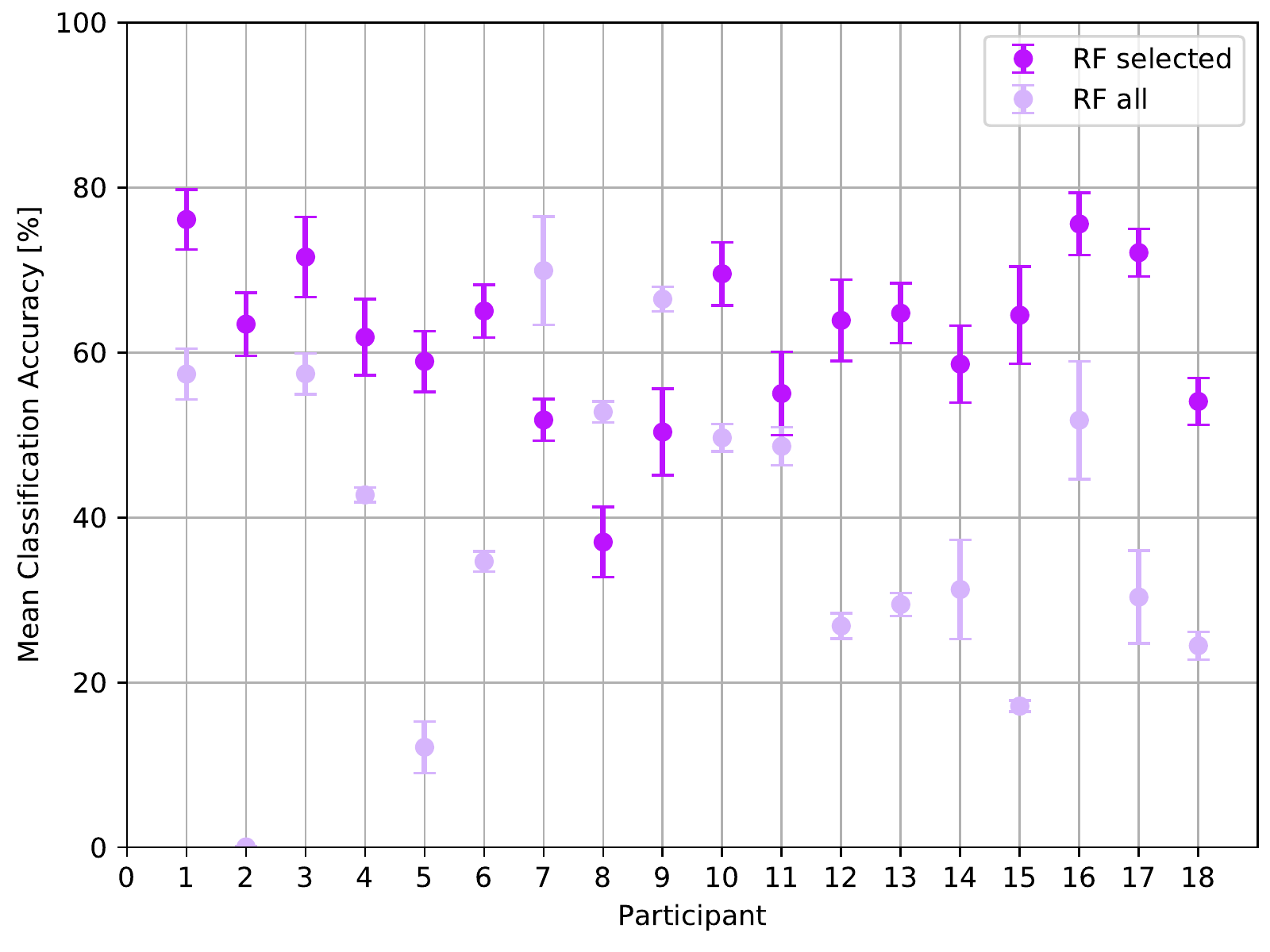}}
    \subfloat[Decision Tree (DT)\label{fig:std_dt}]{%
        \includegraphics[width=0.32\linewidth]{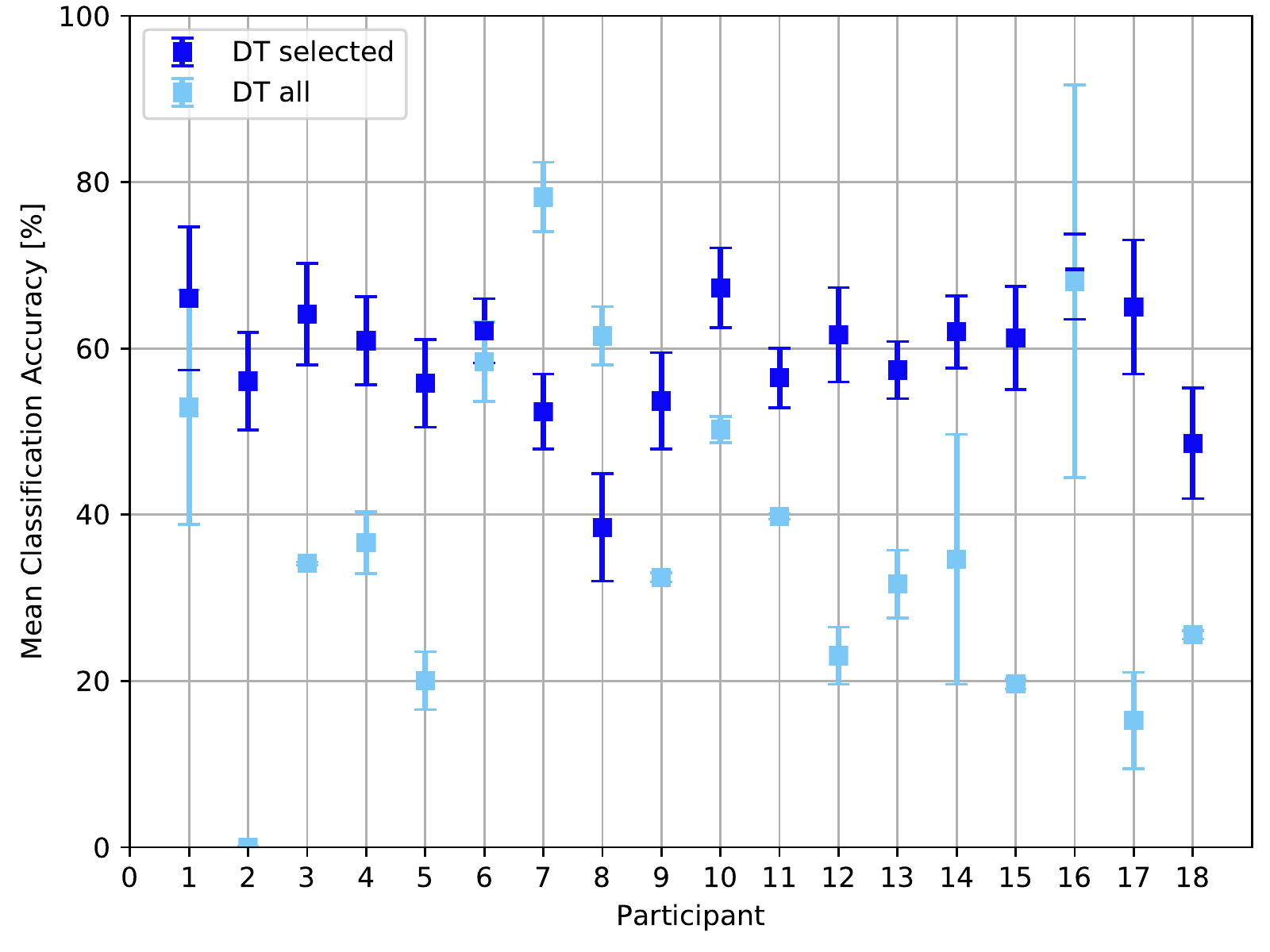}}
    \caption{Mean classification accuracy and standard deviation over all windows for each participant}
    \label{fig:std_classifiers} 
\end{figure*}

In addition to the general classification performance, we were interested in the performance considering single emotion categories. Fig.~\ref{fig:mean_fmeasure} shows the mean f-measure across all participants and window sizes for the selected feature set. We observe that all emotion categories were fairly recognized by the classifiers, with mean f-measures ranging from $0.49$ - $0.79$. In particular, we note that the classifiers achieved higher performances recognizing the~\textit{NEUTRAL} and~\textit{HNHA} emotion categories. Further analysis showed that high arousal categories of emotions were more difficult to distinguish. In case of miss-classification, we observed that the \textit{HPHA} category was incorrectly classified as \textit{NEUTRAL}. However, this occurred less often than the miss-classification with \textit{HNHA}. Also, \textit{NEUTRAL} was rarely miss-classified as \textit{HPHA} or \textit{HNHA}. Consequently, the f-measures of \textit{HPHA}, tend to be lower than \textit{NEUTRAL} and \textit{HNHA}.

Considering the second research question, we investigated the impacts of different window sizes on the classification performance. Fig.~\ref{fig:std_classifiers} depicts the mean classification accuracy and standard deviation over all windows for each participant. We observe that the standard deviations are different for each classifier and participant. In this regard, the KNN shows the lowest standard deviation for all participants, followed by the RF. Also, we note that the all classifiers performed better on the selected feature set than on all features.

\begin{figure}[ht!]
    \centering
    \resizebox{\columnwidth}{!}{
        \includegraphics{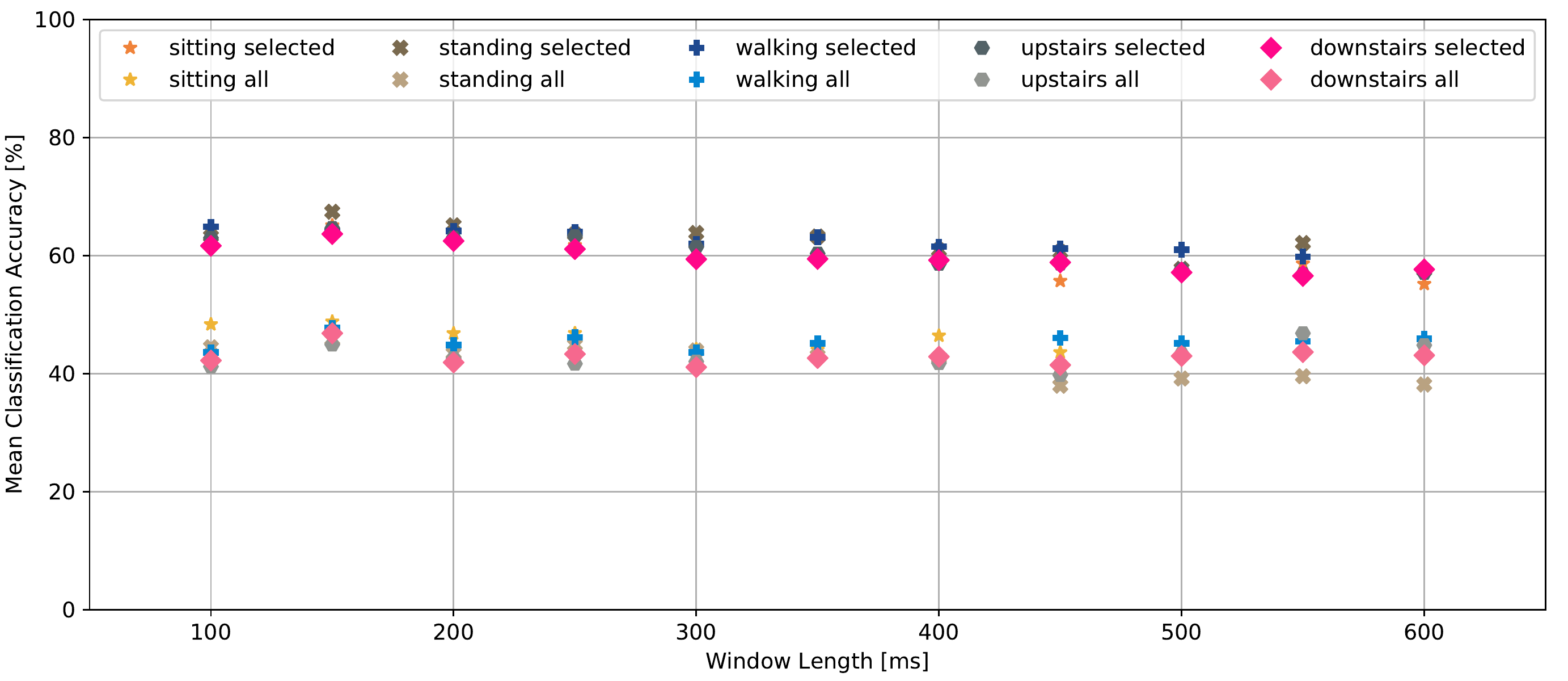}}
    \caption{Mean classification accuracy over all participants for single activities using K-Nearest Neighbors (KNN)}
    \label{fig:mean_activities}
\end{figure}

We also extracted all single activities from scenario~\textit{S-EA} to investigate the accuracy of our models for each activity in the testing phase individually. Note that the models were only trained with data from scenario \textit{S-E} influenced by a low exhausting activity. Fig.~\ref{fig:mean_activities} shows the mean classification accuracy over all participants and for single activities using the KNN classifier. Analog to the previous results, we note that the set with selected features achieved higher classification accuracies than the set containing all features. Considering the selected feature set, emotions were recognized for all activities, ranging from $55.17\%$ - $67.41\%$. For the RF and DT we found similar results ranging from $53.81\%$ to $70.99\%$, and between $50.70\%$ and $65.27\%$, respectively.
\section{Discussion}
\label{discussion}
Considering the aforementioned research questions, we were first interested whether machine-learning models could be trained independently to physiological interferences caused by physical activities~(\textit{RQ1}).

To answer this question, we chose to train emotion models on physiological data influenced by the least exhaustive activity (\textit{S-E}) and tested the performance against the remaining, more exhausting activities (\textit{S-EA}). Overall, our results indicate that the three emotion categories, \textit{NEUTRAL}, \textit{HPHA}, and \textit{HNHA}, can be recognized, ranging from $56.09\% - 66.65\%$ classification accuracy for selected feature sets over all window sizes and per participant. The \textit{NEUTRAL} category achieved the highest f-measure followed by \textit{HNHA} using the RF and DT classifier. An exception was the KNN classifier which achieved higher classification accuracies on the \textit{HNHA} category than the \textit{NEUTRAL} category. However, we noticed that the high arousal emotions, i.e., \textit{HPHA}, \textit{HNHA} were confused with another for all participants. The reason for this might be that the features corresponding to emotion categories of being high arousal are similar in their physiological signal responses. \textit{NEUTRAL} is more often confused with \textit{HPHA} than with \textit{HNHA}. A conceivable cause might be the consequences of selecting the sound samples for \textit{HPHA} and \textit{NEUTRAL} categories, which are closer together on the pleasure scale than on \textit{HNHA}. This decision was made to have two minutes of sounds available for each category.

Regarding the robustness of non-sensitive emotion recognition models~(\textit{RQ2}), we noticed that the size of the sliding window, in general, did not have a significant effect on the classification accuracy. Considering the selected feature, standard deviations range over all participants from $1.64\% - 5.1\%$ for the KNN and $2.52\% -5.89\%$ for RF, and $3.43\% - 8.59\%$ for DT classifier. Nonetheless, we note that standard deviations differ among participants for all classifiers. For example, a significant impact of different window sizes on the classification accuracy was observed for the DT using all features, ranging from $19.85\%$ to $90.70\%$ mean classification accuracy for participant $16$. Whereas participant $11$ is barely influenced by the window size with a standard deviation of 0.34\% using all features. We assume that some window lengths contain more information further used by the classifiers to distinguish between emotion categories. The reason for this might be that emotions and the influence of physical activities on physiological signals are subject dependent due to personal characteristics or the individuality in the execution of physical activities.
\section{Conclusion}
\label{conclusion}
In this paper, we investigated the influence of physical activities, in particular, sitting, standing, walking, walking upstairs and downstairs, on physiological-based emotion recognition. We focused on emotion recognition models, non-sensitive to influences of physical activities. To force non-sensitivity, we trained models with the least exhaustive physical activity (sitting) and tested with the remaining, more exhausting activities. Through our experiment with $18$ participants, we found that non-sensitive models achieved a classification accuracy between $47.88\%$ and $73.35\%$ for three different emotion categories on selected feature sets and per participant. When testing against single activities, a mean classification accuracy between $55.17\%$ and $67.41\%$ was achieved.
Furthermore, we found that features based on the linear regression significantly improved the classification. The relative improvement using only selected features was approximately $20\%$ over using all features. To assess the robustness of non-sensitive emotion recognition models, we investigated different window settings. In this regard, we found no significant influence of different window lengths on the classification performance in general. 

Although the results indicate that emotion recognition models non-sensitive to physiological interferences caused by physical activities are feasible, some research avenues remain. First, more participants as well as more physical activities, including more exhaustive ones, should be investigated to generalize the research findings. In particular, more exhausting activities would cause interferences to a greater extent, probably leading to new perspectives and approaches for such models. Second, our results show that similar emotion categories lying on the same dimensions are challenging to distinguish. In this regard, further research with regards to new features and preprocessing techniques should be targeted.

\ifCLASSOPTIONcompsoc
  \section*{Acknowledgments}
\else
  \section*{Acknowledgment}
\fi

The authors would like to thank all the participants of our experiment for their support.

\bibliographystyle{IEEEtran}

\bibliography{references}
\end{document}